\documentclass[12pt, preprint]{aastex}
\newcommand\BV{\omega_{\rm BV}}
\newcommand\orb{\omega_{\rm orb}}

\newcommand\Edot{{\dot E}_k}

\newcommand\Mg{{M_{11}}}
\newcommand\Msun{{\rm\,M_\odot}}

\newcommand\ergs{{\;\rm erg\; s^{-1}}}
\newcommand\erg{{\;\rm erg}}
\newcommand\kms{{\;\rm km\; s^{-1}}}

\newcommand\kpc{{\;\rm kpc}}
\newcommand\Gyr{{\;\rm Gyr}}
\newcommand\simgt{\lower.5ex\hbox{$\; \buildrel > \over \sim \;$}}
\newcommand\simlt{\lower.5ex\hbox{$\; \buildrel < \over \sim \;$}}

\shorttitle{ICM HEATING AND TURBULENCE BY GALAXIES}
\shortauthors{KIM}
\begin{document}

\title{Heating and Turbulence Driving by 
Galaxy Motions in Galaxy Clusters}

\author{Woong-Tae Kim}
\affil{Department of Physics \& Astronomy, FPRD,
Seoul National University, Seoul 151-742, Republic of Korea}
\email{wkim@astro.snu.ac.kr}

\begin{abstract}
Using three-dimensional hydrodynamic simulations,
we investigate heating and turbulence driving in an intracluster
medium (ICM) by orbital motions of galaxies in a galaxy cluster. 
We consider $N_g$ member galaxies on isothermal and isotropic 
orbits through an ICM typical of rich clusters.  
An introduction of the galaxies immediately produces
gravitational wakes, providing perturbations that
can potentially grow via resonant interaction with the background gas.
When $N_g^{1/2}M_{11} \simlt 100$, 
where $\Mg$ is each galaxy mass in units of $10^{11}\Msun$, 
the perturbations are in the 
linear regime and the resonant excitation of gravity waves is efficient to
generate kinetic energy in the ICM, resulting in 
the velocity dispersion $\sigma_v \sim 2.2 N_g^{1/2}\Mg\kms$.
When $N_g^{1/2}M_{11} \simgt 100$, on the other hand,
nonlinear fluctuations of the background ICM destroy galaxy wakes 
and thus render resonant excitation weak or absent.  In this case, 
the kinetic energy saturates at the level 
corresponding to $\sigma_v \sim220\kms$. 
The angle-averaged velocity power spectra 
of turbulence driven in our models have slopes in the range of
$-3.7$ to $-4.3$. 
With the nonlinear saturation of resonant excitation,
none of the cooling models considered are able to halt cooling catastrophe,
suggesting that the galaxy motions {\it alone} are
unlikely to solve the cooling flow problem.
\end{abstract}
\keywords{cooling flows --- galaxies: clusters : general --- 
turbulence --- waves --- X-rays: galaxies} 

\section{INTRODUCTION}

A lack of cold gas in the central parts of
rich galaxy clusters, as revealed by high-resolution X-ray observations,
has posed a ``cooling flow'' problem,
requiring sources of heat to balance radiative cooling 
of an intracluster medium (ICM). 
Among the proposed heating mechanisms (see \citealt{pet06} for review),
energy injection from active galactic nuclei appears to be the most
favorable, although it requires a central black hole to be more massive than 
observed \citep{fuj04} and is unable to maintain a long term energy 
balance if jets are narrow \citep{ver06}.
Diffusive heating via thermal conduction and/or turbulent mixing
may also be effective if the relevant diffusion coefficient
is quite large and fine tuned (e.g., \citealt{kim03,voi04,den05}).

Less well recognized is the ICM heating by cluster galaxies that 
possess a lot of available energy in their motions
(e.g., \citealt{mil86,bre89}).  
\citet[hereafter BS]{bal90} showed that resonant excitations of internal 
waves driven by orbiting galaxies produce heat at cluster centers 
comparable to radiative loss, provided the galaxy mass is large enough. 
Using Monte-Carlo approaches, \citet{elz04} showed
that dynamical friction of galaxies can be a distributed source of 
heat if the mass-to-light ratio of galaxies exceeds 10.  
\citet{kim05} found that heating by dynamical friction 
reduces the growth rates of thermal instability. 
All these works suggest that the effects of galaxy motions on
thermodynamic evolution of ICM are by no means negligible.

Most of the studies cited above are based on the assumption that the energy 
lost by galaxies is all transferred to {\it thermal} energy of the ICM 
{\it locally} near the galaxies.  This is apparently not
the case because galaxy wakes are spatially extended, overlap with
each other, and induce 
kinetic as well as thermal energies \citep{dei96,ost99}.
The possibility of the ICM stirring by galaxy motions is 
interesting because turbulence appears to be pervasive in 
the ICM (e.g., \citealt{sch04}) and perhaps determines
the characteristic strengths and scales of
cluster magnetic fields (e.g., \citealt{cla01,sub06}).
Although BS allowed for the spatial propagation of internal waves, 
they focused on linear gaseous responses in the WKB limit.
\citet[hereafter LBH]{luf95} explored nonlinear 
evolution of gravity waves, but their models considered a single galaxy
on a radial orbit.  In this Letter, we extend LHB by 
considering a number of cluster galaxies on isotropic orbits.
By varying the mass and number of galaxies, we quantify the 
thermal and kinetic energies induced by galaxy motions and explore
the level and shape of such driven ICM turbulence.

\section{Model and Method}

We consider a galaxy cluster in which the ICM is initially in 
hydrostatic equilibrium under a dark matter potential $\Phi_{\rm DM}$. 
For the initial temperature of the ICM, we adopt a simple form
$T(r) = 7 [1 - 0.6 e^{-(r/r_c)^2}]$ keV with the cooling radius 
$r_c=125$ kpc.  We represent a rigid dark halo using an
NFW profile with the characteristic mass $M_0=6 \times 10^{14} \Msun$ and 
scale radius $r_s=460$ kpc \citep{nav97}.  Although we do not allow for
the presence of a central dominant galaxy considered in LBH,
the cooling core inside $r_c$ in our model still satisfies
the condition for resonant excitation of gravity waves (see \S\ref{reso}).

To study the responses of the ICM to member galaxies, we solve
the ideal hydrodynamic equations:
\begin{equation}
(\partial/\partial t  + \mathbf{v}\cdot\nabla ) \rho + \rho\nabla\cdot 
\mathbf{v}=0,
\end{equation}
\begin{equation}
\rho(\partial/\partial t  + \mathbf{v}\cdot\nabla )  \mathbf{v} 
= -\nabla P -\rho\nabla(\Phi_{\rm DM} + \Phi_g),
\end{equation}
\begin{equation}
\rho(\partial/\partial t  + \mathbf{v}\cdot\nabla ) (e/\rho) 
=-P \nabla\cdot \mathbf{v} - \Lambda,
\end{equation}
where $\Phi_g$ is the time-varying gravitational potential due to
the galaxies and $\Lambda$ is the volumetric cooling rate.
The other symbols have their usual meanings.
The effects of gaseous self-gravity and magnetic fields are ignored.
We adopt an ideal gas law $P=(\gamma-1)e$ with $\gamma=5/3$.
We run both adiabatic (with $\Lambda=0$) and cooling 
(with $\Lambda \neq0$) models by taking 
the cooling function $\Lambda$ used in \citet{rus02}. 

We consider a total of $N_g$ member galaxies distributed within 1 Mpc.
We assume that they follow isotropic and isothermal orbits 
with velocity dispersion $\sigma_r=800$ km s$^{-1}$,
in which case the equilibrium galaxy number density is
$\propto (1+r/r_s)^{\eta r_s/r}$, where $\eta\equiv
2GM_0/(r_s\sigma_r^2)\approx17.5$ \citep{kim05}. 
Under this distribution, about 13\% of the galaxies are located within
$r_c$ at any given time.
We ignore back reaction of the ICM to the galaxies since the energy lost 
due to dynamical friction is small (LBH).
We represent each galaxy using
a Plummer potential $\Phi_p(r)=-GM_g (r^2 + a^2)^{-1/2} $ 
with mass $M_g$ and scale radius $a=7$ kpc; the total perturbing
potential is constructed as 
$\Phi_g (\mathbf r,t) = \sum_{i=1}^{N_g} \Phi_p (|\mathbf r - \mathbf r_i(t)|)$,
where $\mathbf r_i$ is the position vector of the $i$-th galaxy
at time $t$.
To simulate diverse cluster conditions, we vary the number and mass
of the galaxies in the ranges of $N_g\sim 10^2-10^3$ and
$M_g \sim 10^{11}-10^{12} \Msun$.  We report in this work the
results only for models where all the galaxies have equal
masses\footnote{By running a few models in which galaxy masses
vary according to the Schechter function with 
a mass-to-light ratio $M/L\sim L^{0.3}$ \citep{ger01}, we have
confirmed that results are almost unchanged if the breaking
mass $M_*$ in the mass distribution is equal to $M_g$ in the
corresponding fixed mass case.}.

We follow the nonlinear evolution of the hot ICM using a modified
version of the ZEUS code \citep{sto92}, parallelized on a distributed-memory
platform.  Our simulation domain is a cubic box with each side of 2 Mpc;
the center of the box is located at the cluster center. 
We construct a logarithmically spaced Cartesian grid with 256$^3$ zones,
with outflow conditions at all boundaries\footnote{We have also run models 
with 128$^3$ zones, and checked that the results are within less than 10\%
of those from 256$^3$ runs. This suggests 
the energy dissipation caused by numerical diffusion is tolerable.
See \S\ref{heat}.}.  
The grid spacing is 
0.3, 4, and  33 kpc at the center, the cooling radius, and the edge 
of the box, respectively.  
We have confirmed the accuracy of the code by comparing the test results 
for wakes produced by linear-trajectory perturbers with the analytic 
formula of \citet{ost99}.

\section{Nonlinear Simulations}

\subsection{Resonant Excitation by a Single Galaxy}\label{reso}

We first explore the responses of the adiabatic gas to a single galaxy 
moving either on a radial or a circular trajectory. 
BS showed that gravity waves excited by 
a galaxy with orbital frequency $\orb$ become trapped and amplify in 
the region where the local Brunt-V\"ais\"al\"a frequency $\BV$ exceeds $\orb$.
This finding was subsequently confirmed by LBH who ran numerical
simulations for a radial-orbit galaxy.  Our aim here
is to find the dependency of energy injection rate 
on the galaxy mass as well as on the shape of its orbit. 

\begin{figure}
\plotone{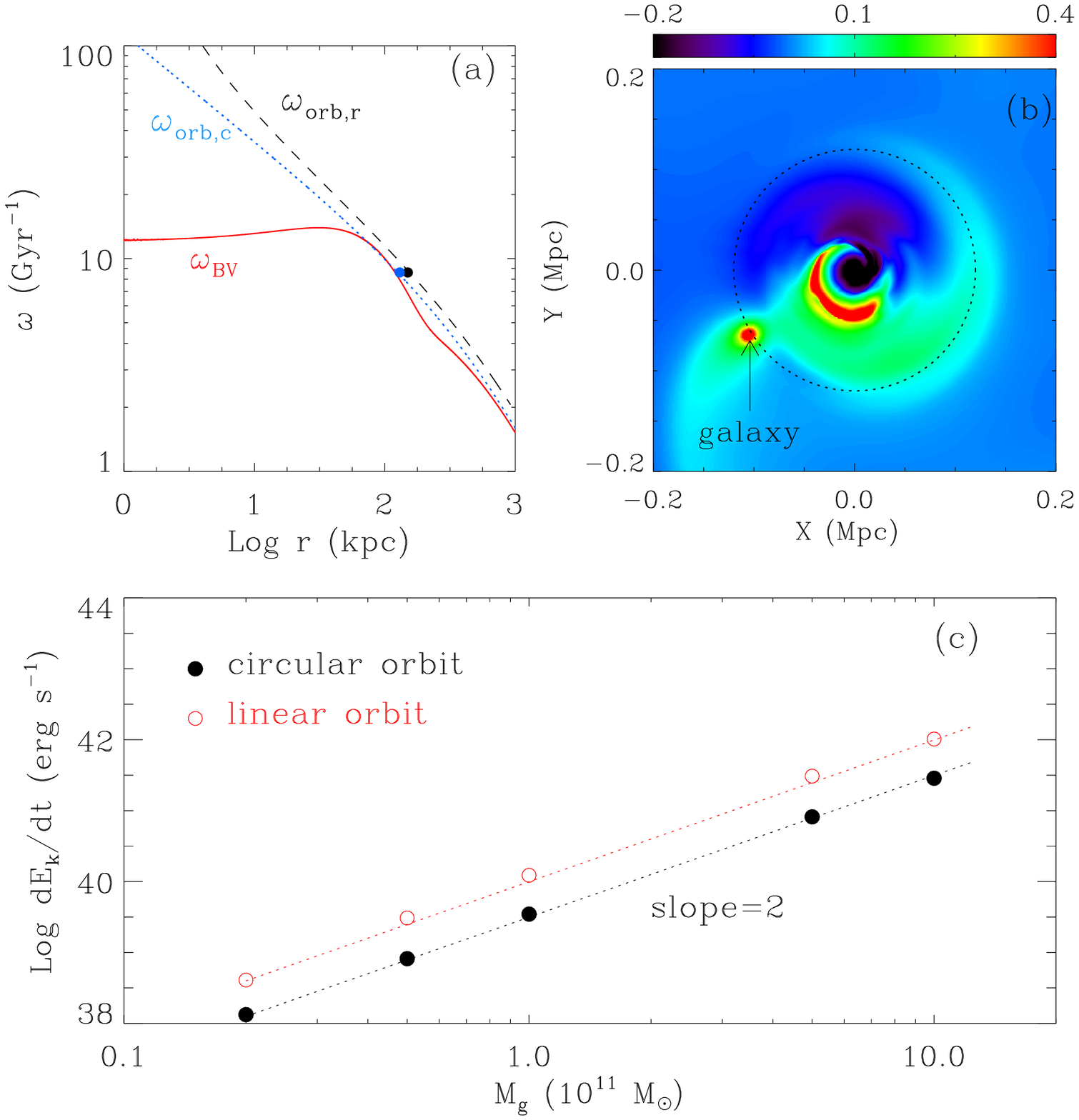}
\caption{(a) Profiles of the radial orbit ($\omega_{\rm orb,r}$), 
circular orbit ($\omega_{\rm orb,c}$), and Brunt-V\"ais\"al\"a 
($\BV$) frequencies in our cluster model. 
(b) $X$-$y$ plane snapshot at $t=6\Gyr$ of the perturbed density 
relative to the initial value in logarithmic scale 
for a circular-orbit galaxy.
(c) Averaged input rates of the kinetic energy due to a single galaxy 
as functions of its mass. 
\label{fig_lbh}}
\end{figure}

Figure \ref{fig_lbh}a shows the radial distributions of 
the radial orbit, circular orbit, and Brunt-V\"ais\"al\"a frequencies 
in our ICM model.  
Even without a central massive galaxy, $\BV$ remains almost flat inside
$80$ kpc, allowing resonant excitation of gravity waves there.  
For models with a radial orbit, 
the galaxy initially set to move with velocity $v=1740\kms$ from
the origin has an average speed of $820\kms$ until it reaches a
turning point at $r=150\kpc$ (black dot in Fig.\ \ref{fig_lbh}a). 
For circular-orbit models, the orbital speed is $v=1060\kms$ 
at $r=130$ kpc (blue dot).  
In both cases, the orbital
period of the galaxy is $0.73\Gyr$, which is chosen to
ensure $\BV > \orb$ inside $r_c$.

As the galaxy starts to move, it generates density and velocity 
perturbations, forming a gravitational wake and imparting some of its 
gravitational energy to thermal and kinetic energies of the ICM.  
In general, the perturbations are a superposition 
of $p$- and $g$-waves.  While high-frequency $p$-waves propagate 
out through a stratified background, 
outgoing $g$-waves reflected at, and remain trapped within, the 
resonance radius ($\sim r_c$ in our models).
The gas near the center receives periodic kicks from the galaxy,
enhancing the levels of density and velocity fluctuations.
Overall evolution of the models with a galaxy on the radial 
orbit is similar to that presented in LBH.

Figure \ref{fig_lbh}b shows a
snapshot of the perturbed density in the orbital plane at $t=6\Gyr$ 
for a model where a galaxy with mass $\Mg\equiv M_g/(10^{11}\Msun)=5$ 
orbits circularly at a near transonic speed (Mach number = 0.97) 
in the clockwise direction.  
It is apparent that perturbations periodically provided by 
the orbiting galaxy is focused to the central part.
The associated vorticity is also well contained inside the resonance
marked by a dotted circle, indicative of 
resonant excitation (LHB).  The density perturbations near the
galaxy in a weak trailing shape are a characteristic feature of a
wake for a circular-orbit perturber \citep{kim07}.

As the galaxy orbits the cluster center and resonantly interacts with
the background gas, the kinetic energy absorbed in the ICM secularly 
increases (approximately linearly) with time,  while showing some
temporal fluctuations.  
We run a number of models with varying $M_g$, and measure the 
rate $\Edot$ at which kinetic energy increases inside $r_c$.
Figure \ref{fig_lbh}c plots the resulting $\Edot$, which are 
fairly well fitted by $\Edot=1.2\times10^{40}M_{11}^2\ergs$ 
and $\Edot=3.5\times10^{39}M_{11}^2\ergs$ 
for the radial- and circular-orbit cases, respectively.
For the model parameters we adopt, therefore, a galaxy on a radial orbit
is about three times more efficient in driving kinetic energy into the 
ICM than the circular-orbit counterpart, since it in the former orbit
can traverse the central region directly (BS).
The dependency of $\Edot$ on $M_g^2$ indicates that 
perturbations in all the models with a single galaxy 
are in the linear regime.

\subsection{Heating by Cluster Galaxies}\label{heat}

\begin{figure}
\plotone{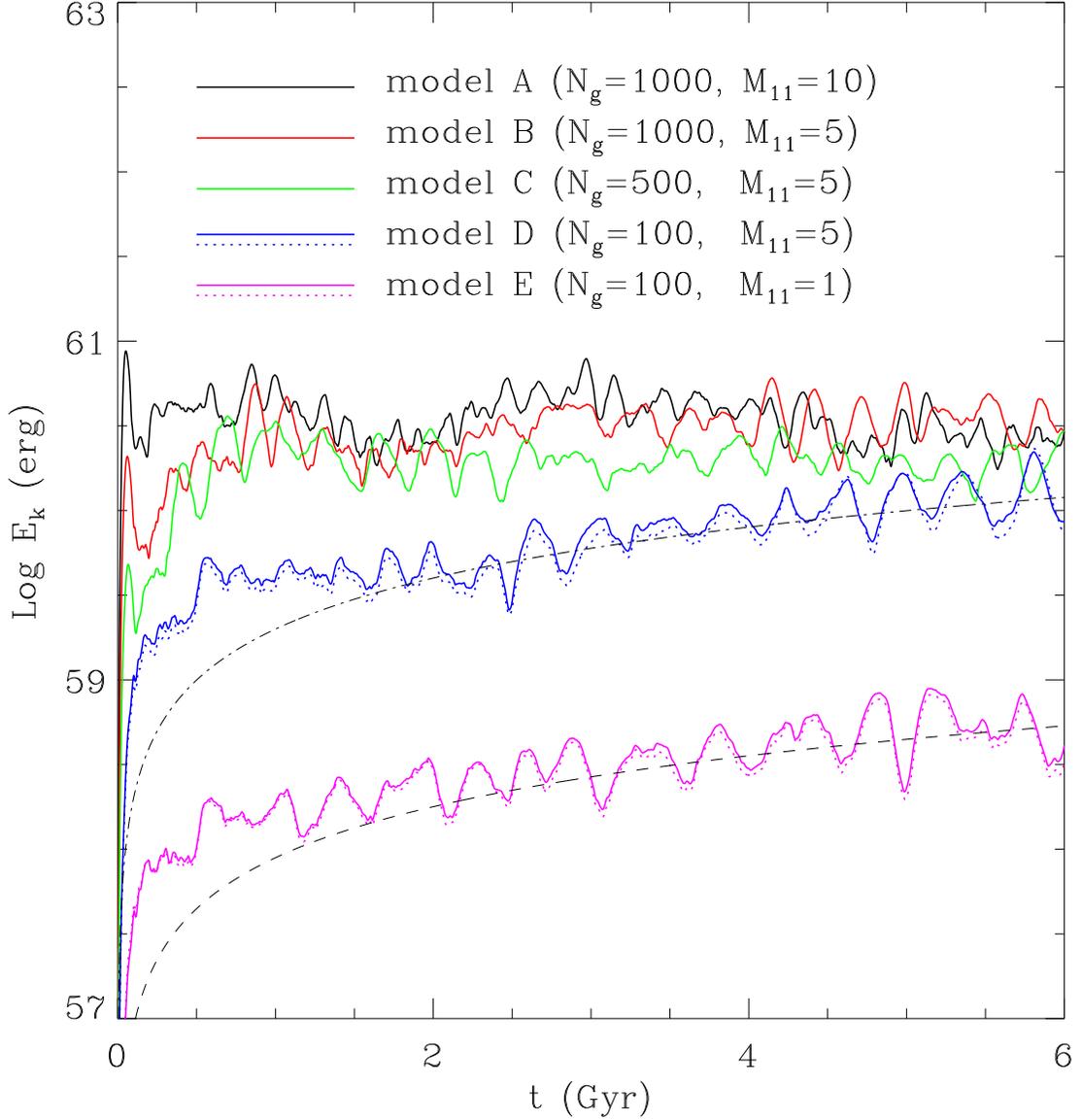}
\caption{Temporal evolution of the kinetic energy $E_k$ inside the cooling
radius for adiabatic models that differ in the number and mass of 
the galaxies.  Solid curves show the results from runs with $256^3$ zones.
Dotted lines (for models D and E) from $128^3$-zone runs are
within 8\% of the corrresponding solid curves.
Dashed and dot-dashed lines draw 
$E_k = 8.2\times 10^{57}M_{11}^{2}(t/\Gyr)\erg$ for $\Mg=1$ and 5,
respectively.
\label{fig_ekdot}}
\end{figure}

We now consider more realistic cluster models in which the ICM is 
continuously stirred by many member galaxies.  We run nine adiabatic 
models as well as nine cooling counterparts.
Figure \ref{fig_ekdot} plots time evolution of the kinetic energy $E_k$ of
the ICM inside $r_c$ for some of the adiabatic models.
A sudden introduction of the galaxies causes the ICM to respond abruptly,
initiating an rapid increase of $E_k$ for $t<0.1\Gyr$.
For models D ($N_g=10^2$, $\Mg=5$) and E ($N_g=10^2$, $\Mg=1$), the
perturbations are initially in the linear regime and soon begin to  
interact resonantly with the background gas.
As gravity waves concentrate toward the center and amplify, $E_k$
grows secularly with time.
For $t>2\Gyr$, $E_k$ in models D and E 
(and other models with $N_g^{1/2}M_{11} \simlt 100$) increases
almost linearly at a rate ${\dot E}_k=2.6\times 10^{41}(N_g/10^2)M_{11}^{2}
\ergs$.  Compared to the results of \S\ref{reso}, this corresponds to
about 20 radial-orbit galaxies taking part in resonant excitation 
of the ICM inside $r_c$. On the other hand, 
in model A ($N_g=10^3$, $\Mg=10$) where initial perturbations 
are very strong,  $E_k$ stays almost 
constant at $\sim10^{60.5}\erg$ during its entire evolution,
indicating weak or no resonant excitation. 
Models B ($N_g=10^3, \Mg=5$) and C ($N_g=500, \Mg=5$) that 
have weaker initial perturbations than model A 
enhance $E_k$ to the saturation level $\sim 10^{60.5}\erg$
for $t\simlt 1\Gyr$, after which $E_k$ again remains constant. 
At saturation, the amplitudes of density fluctuations 
are about 40\% relative to the mean value, easily disrupting galaxy
wakes that would otherwise supply fresh perturbations for resonant 
excitation. 
Consequently, resonant excitation becomes weak or even absent 
in a highly nonlinear background.

Unlike kinetic energy that grows with time when $N_gM_g^2$ is sufficiently 
small, we find that thermal energy of the ICM does not show a clear 
indication of resonant excitation.
The total amount of thermal energy within $r_c$ 
driven by a mere introduction of the galaxies is  
$\Delta E_t = 10^{58}N_g\Mg\erg$, which is already large and stays 
more or less constant with time for all the models.  This may be because 
kinetic energy is more prone to resonant excitation, or because 
the amount of heat supplied (presumably at a rate similar to $\Edot$) 
by resonant excitation is much smaller than $\Delta E_t$, so that
it does not readily manifest in the energy curves over time. 
The ratio of kinetic to thermal energies in a wake produced by a 
single galaxy with size $a$ and velocity $\sigma_v$ 
is roughly $\sim(GM_g/\sigma_v a)^2/(3c_s^2)$, where $c_s$ is the 
adiabatic sound speed of the ICM \citep{jus90}.
Since this ratio is 
less than $10^{-2}$ when $\Mg\leq1$ for typical parameters we adopt, 
kinetic energy is certainly a much better tracer
of resonant excitation.

To check if galaxy motions and the associated heating can solve the 
cooling flow problem, we repeat the simulations by
including the radiative cooling explicitly.
Our models lose thermal energy at a rate $L_X\sim10^{44}\ergs$ 
from the cooling core. 
Without any heat source, they would experience 
a cooling catastrophe within $0.6\Gyr$. 
It turned out that none of the models we considered were able to 
prevent, albeit considerably delay, the runaway cooling.
For instance, the cooling model with 
$N_g=10^2$ and $\Mg=5$ (corresponding to model D)
undergoes a catastrophic event at $2.5\Gyr$, 
while the model  $N_g=10^3$ and $\Mg=10$
(corresponding to model A) develops a cooling flow in $1.7\Gyr$ near the
center.  The heating rate due to 
resonant excitation is probably lower in the latter model, 
for which the perturbed kinetic energy saturates immediately.

\subsection{Properties of ICM Turbulence}

\begin{figure}
\plotone{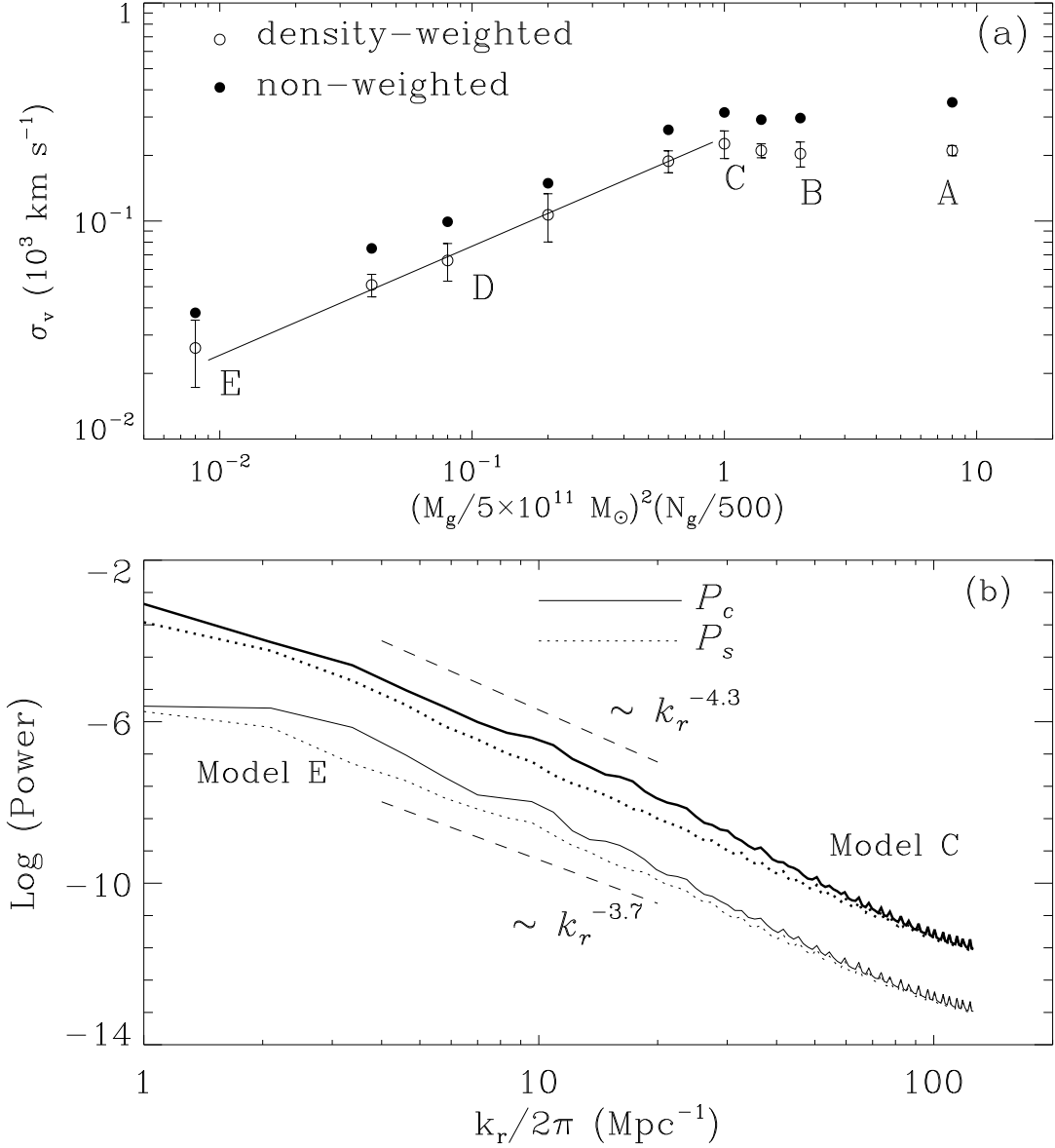}
\vspace{-0.2cm}
\caption{(a) Three-dimensional velocity dispersions $\sigma_v$ 
of the ICM turbulence driven by 
galaxy motions as functions of $N_gM_{11}^2$.  Open circles, with errorbars 
representing the standard deviations in the temporal fluctuations 
of $\sigma_v$, denote the density-weighted values, while the non-weighted 
ones are given by filled circles.
The solid line corresponds to $\sigma_v = 2.2 \kms N_g^{1/2}\Mg$.  
(b) Spherically binned power spectra of (solid line) compressive and
(dotted line) shearing parts of the ICM velocity in models C and E
at $t=6\Gyr$.
\label{fig_turb}}
\end{figure}

We have seen that motions of the member galaxies generate 
a large amount of kinetic energy in the ICM.
The kinetic energy is in the form of fluctuating isotropic velocity
fields with vanishingly small mean values.  
We regard these spatially-uncorrelated, random gas motions as 
ICM turbulence.  To quantify the turbulence level,
for each run we measure the velocity dispersions of the gas inside $r_c$.
Figure \ref{fig_turb}a plots as open circles the resulting 
density-weighted 3D velocity dispersion  $\sigma_v\equiv
(2E_k/M_c)^{1/2}$ averaged over $3-6\Gyr$ as a function of $N_gM_{11}^2$, 
where $M_c$ is the ICM mass inside $r_c$.  
Filled circles give non-weighed velocity dispersions,
which is larger than the density-weighted values by about a factor of 1.5. 
Consistent with the results of the previous subsection, $\sigma_v$ 
increases linearly with $N_g^{1/2}M_{11}$ 
until $N_g^{1/2}M_{11}\approx 100$, beyond 
which $\sigma_v$ is approximately constant at $\sim210-230\kms$. 

To characterize the turbulence driven by galaxy motions, we calculate
Fourier power spectra of the compressive and shear 
components of ICM velocities defined by
$v_{c}^2 = |\hat{\mathbf k}\cdot {\mathbf v}_k |^2$ and
$v_{s}^2 = |\hat{\mathbf k}\times {\mathbf v}_k |^2$,
respectively. Here, ${\mathbf v}_k$ is the Fourier transformed velocity
and $\hat{\mathbf k}$ is the wavenumber.  We then 
bin them spherically in the $\hat{\mathbf k}$-space and  
calculate the angle-averaged power spectra $P_{c}(k_r)$
and $P_{s}(k_r)$ 
as functions of the radial wavenumber $k_r$.
Figure \ref{fig_turb}b shows $P_{c}$ and
$P_{s}$ for models C and E at $t=6\Gyr$.
The ratio of total power in the compressive to
shearing parts is about 2.5 for both models, indicative of 
fairly subsonic turbulence (e.g., \citealt{ves03}).
The power index of the shearing part is $-3.7$ in the inertial 
range for model E, which becomes steeper with increasing $\sigma_v$,
yielding $-4.3$ for model C\footnote{Caution should be made 
in interpreting the power indices in comparison with Kolmogorov power
spectra, since in our models the background density is not uniform and 
the velocity field is not periodic.}.
The compressive parts have similar slopes, although 
they have excess power at $\sim 50-100$ kpc scales, 
which appear to be associated with the mean galaxy
separations\footnote{The galaxies are concentrated more strongly
toward the center and have a mean 
distance of 82 kpc inside the cooling radius.}.

\section{Discussion}

Galaxies in a cluster contain a plentiful amount of energy 
in their orbital motions that can be transferred to turbulent kinetic 
and thermal energies of the ICM 
(e.g., BS; \citealt{dei96}).  
In this Letter, we have shown that resonant excitation of gravity
waves driven by galaxy motions is efficient only when density
and velocity fluctuations in the 
background are in the linear regime, 
while becoming inefficient when the background medium exhibits 
large amplitude fluctuations.
Although it is uncertain at what rate resonant excitation heats the ICM,
our numerical results suggest that heating by galaxy motions is 
insufficient to quench the cooling catastrophe.  If the heating
rate is similar to the kinetic energy injection rate $\Edot$ we found, the
energy balance between the heating and X-ray cooling 
requires $N_g^{1/2}M_{11} \sim 200$.  
Although rich clusters may 
contain enough number and mass of galaxies to satisfy this condition, 
resonant excitation will switch off when this condition is met.
We thus conclude that heating by galaxy motions {\it alone}
cannot be the main solution to 
the cooling flow problem, although it can delay 
the catastrophic event significantly
(e.g., BS; \citealt{kim05}).

The two key parameters that control the rate of energy injection
and the level of turbulence are the mass and number of galaxies.
Many uncertainties surround the observational determinations of
the average galaxy mass, but a recent analysis using strong-lensing models
shows that a cluster galaxy can have mass as large as 
$5\times 10^{11}\Msun$
including a dark halo \citep{hal07}.  For rich clusters with
$N_g>10^3$ inside 1 Mpc, therefore, our numerical results suggest 
that ICM turbulence driven solely by galaxy motions is 
probably in a saturated state with $\sigma_v \sim 220\kms$.
Given the many arbitrary choices for the cluster parameters, 
this value is in rough agreement with an analytic estimate of $\sim300\kms$
by \citet{sub06} who used scaling laws of hydrodynamic wakes
without considering nonlinear saturation.
Note that the saturated $\sigma_v$ is similar to those required
to explain the observed magnetic field strength ($\sim 1\mu$G) in terms
of energy equipartition (e.g., \citealt{gol91}).

The angle-averaged velocity power spectra of turbulence driven in our models 
are characterized by inertial-range slopes ranging from 
$-3.7$ to $-4.3$.  It is interesting to note that these 
are comparable to the values between $-11/3$ and 
$-13/3$ inferred from the observed pressure maps of
the ICM in the Coma cluster \citep{sch04}, although 
density inhomogeneities created by recent infall/mergers 
(e.g., \citealt{ada05})
are likely to influence the pressure maps 
of the Coma cluster that is dynamically young.
Obviously, there are other potential driving sources including AGN and 
subcluster mergers, ram pressure stripping, etc.
It will be interesting to see how turbulence driven
by each process adds together when nonlinear effects 
as well as magnetic fields are considered.

\acknowledgments

The author is grateful to an anonymous referee for a helpful report.
This work was supported by Korea Science and Engineering
Foundation (KOSEF) grant R01-2004-000-10490-0. 
The numerical computations presented in this work were performed on the
Linux cluster at KASI 
built with funding from KASI and ARCSEC.

\clearpage

\clearpage

\clearpage

\end{document}